\documentstyle[12pt]{article}
\begin{document}
\begin{flushright}
{\large PRL-TH-94/7}
\vskip .2cm
{\large UNIL-TP-2/94}
\vskip .2cm
{\large March 1994}\\
\end{flushright}
\vskip .5cm
\begin{center}
{\Large \bf Effect of longitudinal electron polarization \\ in the
measurement of the $\tau$ weak dipole moment \\
in $e^+e^-$ collisions }\\[50pt]
{\bf B. Ananthanarayan}
\vskip .2in Institut de Physique Th\'{eorique},
B\^{a}timent des Sciences Physiques,\\
Universit\'{e} de Lausanne,
CH 1015, Lausanne, Switzerland.\\
\vskip .2in
and
\vskip .2in
{\bf Saurabh D. Rindani}\\ \vskip .2in Theory Group,
Physical Research Laboratory\\Navrangpura, Ahmedabad 380009, India\\[40pt]
{\bf \underline{Abstract}}\\[20pt]
\end{center}

We show that certain CP-odd momentum correlations in
the production and subsequent decay of tau pairs in $e^+e^-$
collisions get enhanced when the electron beam is longitudinally
polarized.  Analytic expressions for these correlations are
obtained when
$\tau^+\tau^-$ have a  ``weak" dipole form factor (WDFF)
coupling to $Z$ in the case of
two-body decay modes of the $\tau$. Expressions of the
variance of these correlations due to the standard
model interactions are also presented.
For $e^+e^-$ collisions at the $Z$
peak, a sensitivity of about $10^{-17}
e$ cm for the real part of the
$\tau$ WDFF and $3\times 10^{-17}
e$ cm for the imaginary part of the $\tau$ WDFF
can be reached using $\pi$ and $\rho$ channels
(which are the most sensitive channels) in $\tau$ decay,
with $10^6\, Z$'s likely to be available at the SLC
at Stanford with $e^-$ polarization of 62\%-75\%.
\newpage

\noindent{\bf 1. Introduction}
\vskip 1cm

The study of CP-odd correlations has been proposed in the past as
a test of CP [1]. The observation of such correlations, for example
in $e^+e^-$ or $p\bar{p}$ experiments, would signal violation of
CP.  In particular, CP-odd observables arising due to a possible
$\tau$-lepton electric or ``weak" dipole moment have been analyzed
in great detail [2,3].  Such dipole moments arise in
extensions of the standard model (SM) with CP violation coming
necessarily from sectors other
than the standard $3\times 3$ Cabibbo-Kobayashi-Maskawa matrix
in order to be numerically significant.

Recent experiments at the LEP collider in CERN  have also put an
experimental upper limit [4] on the real part of the weak dipole
form factor (WDFF) of $\tau$ at the $Z$ resonance by looking for
the tensor correlation $T_{ij} = ({\bf q}_+ -{\bf q}_-)_i ( {\bf
q}_+\times  {\bf q}_- )_j$, where ${\bf q}_+({\bf q}_-)$
represents the momentum of a charged particle arising from the
decay of $\tau^+(\tau^-)$ produced in $Z$ decay.

Recently, the Stanford Linear Collider (SLC) has achieved a
longitudinal polarization $(P_e)$ of 62\% for the electron beam,
which is likely to be increased to about 75\% [5].  Longitudinal
polarization of $e^-$ of 22\% has already been used for the
determination of $\sin ^2\theta_W$  using left-right asymmetry
[6]. Experiments with the present polarization of 62\% can lead to
better accuracy than obtained so far at LEP [7].

We study here the effect of longitudinal $e^-$ polarization on
CP-odd momentum correlations in $e^+e^- \rightarrow
\tau^+\tau^-$ with the subsequent decay of $\tau^+$ and
$\tau^-$. Since we have mainly the ongoing experiments at SLC in
mind, we concentrate on the $e^+e^-$ centre-of-mass (c.m.)
energy tuned to the $Z$ resonance.
CP violating effects give rise to CP-odd correlations among
$\tau$ momenta and
spins. In practice, tau spin is measured by looking at its decay
products, and we need
consider only the directly observable momentum correlations.
 The CP-odd momentum
correlations are associated with the c.m. momenta
${\bf p}$ of $e^+$,
${\bf q_{\bar B}}$ of $\bar{B}$ and
${\bf q_A}$ of $A$, where the $\bar{B}$ and
$A$ arise in the decays $\tau^+ \rightarrow \bar{B}+ \bar{\nu}_{\tau}$
and $\tau^-\rightarrow A + \nu_{\tau}$, where $A,\ B$ run over
$\pi$, $\rho$, $A_1$, etc.  In the case when A and B are
different, one has to consider also the decays with A and B
interchanged, so as to construct correlations which are
explicitly CP-odd.

This work extends the work of an earlier
paper [8], where we had obtained analytic expressions for the
correlations in the single-pion decay mode of the tau. Besides
including more details, the calculations of [8] are
now generalized to include
other two-body decay modes of the $\tau$ in general
and is applied specifically to the case of $\tau\rightarrow
\pi+\nu_\tau$ and $\tau\rightarrow \rho+\nu_\tau$ due to the
fact that these modes possess a good resolving power
of the $\tau$ polarization, parametrized in terms of
the constant $\alpha$ which takes the value 1 for the
$\pi$ channel (with branching fraction of about
$11\%$) and 0.46 for the $\rho$ channel [9] (with branching
fraction of about $22\%$).
It may be noted that with these final states the
substantive fraction of the channels that are sensitive
to such correlations are accounted for; three body
leptonic final states must also be included; they
are characterized by a somewhat smaller $\alpha=-0.33$
(with branching fraction of
about $35\%$).  Thus with the channels studied here, one
more or less reaches the limits of discovery in
such experiments.
(It would
also be possible to apply this to the decay $\tau\rightarrow
A_1+\nu$;  $\alpha_{A_1}$ is however too small
to be of any experimental relevance.)
Further, we
also present closed form expressions for the variance
of the correlations considered due to standard model
interactions.
 These, because of finite statistics, provide a
measure of the CP-invariant background to the determination of
the CP-odd contributions to the correlations. In case of a
negative result, the limit on the CP-violating interactions is
obtained using the value of the variance and the size of the
data sample.

It must be noted that correlations which are CP violating in the
absence of initial beam polarization are not strictly CP odd
for arbitrary $e^+$ and $e^-$ polarizations, since the initial
state is then not necessarily  CP even. We argue, however,
that this is true to a high degree of accuracy in the case at
hand.

Our main result is that certain CP-odd correlations, which are
relatively small in the absence of $e^-$ polarization, since they
come with a factor $r=2g_{Ve}g_{Ae}/(g^2_{Ve} +
g^{2}_{Ae})$ ($\approx$ 0.16), get enhanced in the presence of
polarization, now being proportional to
$(r - P_e)/(1-rP_e)$ ($\approx
0.71$ for $P_e = -.62$) [10].  Here $g_{Ve}, g_{Ae}$ are the
vector and axial vector couplings of $e^-$ to $Z$.  The
correlations which have this property are those which have an
odd number of factors of the $e^+$ c.m. momentum
${\bf p}$, since this would need P and C
violation at the electron vertex.  Furthermore, we suggest a
procedure for  obtaining these correlations from the difference in the
event distributions for a certain polarization $P_e$ and the
sign-flipped polarization $-P_e$. With this procedure, the correlations
are further enhanced, leading
to increased sensitivity.
The inclusion of the $\rho$ channel leads to a considerable
improvement in the sensitivity that can be reached in the
measurement of Im-$\tilde{d}_{\tau}$ while improving the
measurement of Re-$\tilde{d}_\tau$ less spectacularly.
It is once again worth noting that these simple vector
correlations can probe both the magnitude and phase of
the WDFF in the presence of large polarization despite the
modest luminosity of SLC to accuracies comparable to (or even
better than)
those of the tensor correlations suggested in [3].

More specifically, we have considered the observables $O_1
\equiv \frac{1}{2} \left[\hat{\bf p}\cdot \left({\bf q}_{\bar B}\times
{\bf q}_A \right)\right.$\linebreak $+\left.\hat{\bf p}\cdot
\left({\bf q}_B\times  {\bf q}_{\bar A} \right)\right]$ and $O_2 \equiv
\frac{1}{2}\left[\hat{\bf p}\cdot \left({\bf q}_A + {\bf
q}_{\bar B} \right)+\hat{\bf p}\cdot \left({\bf q}_{\bar A} +
{\bf q}_B \right)\right]$ (the caret denoting a unit vector) and
obtained analytic
expressions for their mean values and standard deviations in the
presence of longitudinal $e^-$ polarization $P_e$.  By the
procedure outlined above, the
WDFF at the $Z$, $\tilde{d}_{\tau}(m_Z)$
which can be measured at $1\sigma$ level is about 5$\times
10^{-17} e$ cm ($10^{-17} e$ cm) for a sample of 50,000
$(10^6)$ Z's, using these two-body $\tau$ decay channel.
Moreover, $O_2$,
being CPT-odd, measures ${\rm Im}\,\tilde{d}_{\tau}$, whereas $O_1$
measures ${\rm Re}\,\tilde{d}_{\tau}$.
Inclusion of other exclusive $\tau$
decay modes (not studied here) would improve the sensitivity
further.
\vskip 1cm
\newpage
\noindent{\bf 2. Notation and Formalism}

\vskip 1cm

Although much of this section has already been described
in our previous paper [8] we will repeat it for the sake
of completeness and to make the generalization from
the $\pi$ channel case more transparent.

The process we consider is
\begin{equation}
e^-(p_-) + e^+(p_+)\rightarrow \tau^-(k_-) + \tau^+(k_+),
\end{equation}
with the subsequent decays
\begin{equation}
\tau^-(k_-)\rightarrow A(q_A) + \nu_{\tau} ,\;
\tau^+(k_+)\rightarrow \bar{B}(q_{\bar B}) +\overline{\nu}_{\tau} ,
\end{equation}
together with decays corresponding to $A$ and $B$ interchanged
in (2).

Under CP, the various three-momenta transform as
\begin{equation}
{\bf p}_-\leftrightarrow -
{\bf p}_+ ,\;
{\bf k}_-\leftrightarrow -
{\bf k}_+ ,\;  {\bf q}_{A,B}
\leftrightarrow - {\bf q}_{{\bar A},{\bar B}} .
\end{equation}
We choose for our analysis the two CP-odd observables $O_1
\equiv \frac{1}{2} \left[\hat{\bf p}\cdot \left({\bf q}_{\bar B}\times
{\bf q}_A \right)\right.$\linebreak $+\left.\hat{\bf p}\cdot
\left({\bf q}_B\times  {\bf q}_{\bar A} \right)\right]$ and $O_2 \equiv
\frac{1}{2}\left[\hat{\bf p}\cdot \left({\bf q}_A + {\bf
q}_{\bar B} \right)+\hat{\bf p}\cdot \left({\bf q}_{\bar A} +
{\bf q}_B \right)\right]$, which have an odd number of factors
of $\hat{\bf p}$, the unit vector along ${\bf p}_+$.  As
mentioned before, they are expected to get
enhanced in the presence of $e^-$ polarization.

Though these observables are CP
odd, their observation with polarized $e^+$ and $e^-$ beams is
not necessarily an indication of CP violation, unless the $e^+$
and $e^-$ longitudinal polarizations are equal and opposite, so
that the initial state is described by a CP-even density matrix.
The case we consider here, namely, when only the $e^-$ is
polarized, therefore needs further discussion.

In the case when only the electron beam is polarized, the
CP-invariant interactions can in principle contribute to the
correlations we have calculated. However, in the limit of
$m_e=0$, the couplings of like-helicity $e^+e^-$ pairs to spin-1
states like $\gamma$ and $Z$ drop out, effectively giving rise
to a CP-even initial state to a very good accuracy for arbitrary
$e^+$ and $e^-$ polarizations. The exact
extent to which CP-invariant
interactions would contribute to the correlations would be model
dependent, and has to be calculated . However, it is clear that
at the $Z$ resonance, the corrections would be suppressed at
least by a factor $m_e/m_Z \approx 6\times 10^{-6}$.

There are however corrections at order $\alpha$ from tree-level
helicity-flip collinear photon emission which are not suppressed
by electron mass [11,12]. We have not considered any order-$\alpha$
corrections in this work. However, if there is a CP-invariant
background at order $\alpha$ to the correlations we calculate,
this could completely wipe out the small CP-violating  effects
we are concerned about. Fortunately, the helicity-flip cross
section being non-resonant is suppressed by a factor of about
$10^{-4}$ in the region of the resonance [12].
It is therefore expected that the corresponding correlations
will be small enough not to affect our sensitivity estimates,
and can be neglected. (In fact, for the
case of our observable $O_1$ which is T odd, we need not worry
about this, since CP invariant
contributions necessarily need to have an absorptive part, which
is not present in the tree-level order-$\alpha$ correction).

There may be higher-order non-resonant CP-odd
helicity combinations (without collinear photon emission) which
contribute to $\langle
O_{1,2}\rangle$ at one loop. Since these do not interfere with
the leading $Z$ contributions coming from CP-even helicity
combinations, they will be suppressed by a factor $\alpha
^2\Gamma^2_Z/m^2_Z  \approx 5\times 10^{-8}$.

Of $O_1$ and  $O_2$, $O_1$ is even under the combined CPT transformation,
and $O_2$ is CPT-odd. A CPT-odd
observable can only have a non-zero value in the presence of an
absorptive part of the amplitude.  It is therefore expected that
$\langle O_2\rangle $ will be proportional to the imaginary part of the weak
dipole form factor ${\rm Im}\,\tilde{d}_{\tau}$ , since final-state
interaction, which could give rise to an absorptive part,
is negligible in the weak $\tau$ decays. Since $\langle O_1\rangle $ and mean
values of other CPT-even quantities will be proportional to
${\rm Re}\,\tilde{d}_{\tau}$, phase information on
$\tilde{d}_{\tau}$ can only be obtained if $\langle O_2\rangle $ (or
some other CPT-odd quantity) is also measured.

We assume SM couplings for all particles except $\tau$, for which
an additional WDFF interaction is assumed, viz.,
\begin{equation}
{\cal L}_{WDFF} = - \frac{i}{2} \tilde{d}_{\tau}
\overline{\tau}\sigma^{\mu\nu}\gamma_5\tau
\left(\partial_{\mu}Z_{\nu} - \partial_{\nu}Z_\mu\right) ,
\end{equation}
where $\tilde{d}_{\tau} \equiv \tilde{d}_{\tau}
(s\!=\!m^2_Z)$.  Using (4), we now proceed to calculate
$\langle O_1\rangle $ and $\langle O_2\rangle $ in the presence
of longitudial polarization $P_e$ for $e^-$.

We can anticipate the effect of $P_e$ in general for the process
(1).  We can write the matrix element squared for the process in
the leading order in perturbation theory, neglecting the
electron mass, as
\begin{equation}
\vert M\vert^2 = \sum_{i,j} L^{ij}_{\mu\nu}(e) L^{ij
\mu\nu*}(\tau) \frac{1}{s-M^2_i}\, \frac{1}{s-M^2_j} ,
\end{equation}
where the summation is over the gauge bosons $(\gamma ,Z,\ldots)$
exchanged in the $s$ channel, and $L^{ij}_{\mu\nu}(e,\tau)$
represent the tensors arising at the $e$ and $\tau$ vertices:
\begin{equation}
L^{ij}_{\mu\nu} = V^i_\mu V^{j*}_{\nu} .
\end{equation}
For the electron vertex, with only the SM  vector and axial vector
couplings,
\begin{equation}
V^i_{\mu}(e) = g_i\bar{v}(p_+ ,
s_+)\gamma_{\mu}\left(g^i_{Ve}-\gamma_5g^i_{Ae}\right)u(p_- ,
s_-),
\end{equation}
$g_i$ being the appropriate coupling constant $\left( g_{\gamma}
= e, g_Z = g/(2\cos\theta_W)\right)$, and $g^i_{Ve}$ and
$g^i_{Ae}$ are given by
\begin{equation}
g^{\gamma}_{Ve} = - 1,\; g^{\gamma}_{Ae} = 0 ;
\end{equation}
\begin{equation}
g^Z_{Ve} = -\frac{1}{2} + 2\sin ^2\theta_W ,\; g^Z_{Ae} =
-\frac{1}{2} .
\end{equation}
It is easy to check, by putting in helicity projection operators,
that
\begin{eqnarray}
\lefteqn {L^{ij}_{\mu\nu}(e)=g_ig_j} &&\nonumber \\
 &\! \times&\!\!\!\! \left\{\left[\left( 1-P_eP_{\bar
e}\right) \left(g^i_{Ve}g^j_{Ve}
+ g^i_{Ae}g^j_{Ae}\right) - \left( P_e-P_{\bar
e}\right)\left(g^i_{Ve}g^j_{Ae} +
g^i_{Ae}g^j_{Ve}\right) \right]
Tr(\rlap{$p$}/_-\gamma_{\mu}\rlap{$p$}/_+\gamma_\nu) \right.
\nonumber \\
 &\! +&\!\!\!\! \left.  \left[ \left( P_e-P_{\bar e}\right)
\left(g^i_{Ve}g^j_{Ve} + g^i_{Ae}g^j_{Ae}\right) -
\left( 1-P_eP_{\bar e}\right)\left(g^i_{Ve}g^j_{Ae} +
g^i_{Ae}g^j_{Ve}\right)\right]
Tr \left(\gamma_5\rlap{$p$}/_-
\gamma_{\mu}\rlap{$p$}/_+\gamma_\nu \right) \right\} \nonumber\\
&&
\end{eqnarray}
in the limit of vanishing electron mass, where $P_e$ ($P_{\bar
e}$) is the
degree of the $e^-$ ($e^+$) longitudinal polarization. Note that
the combinations $P_e-P_{\bar e}$ and $1-P_eP_{\bar e}$ occurring
in (10) are indeed CP-even, showing that the initial state is
effectively CP even for arbitrary $P_e$, $P_{\bar e}$.

Eq.(10) gives a simple way of incorporating the effect of the
longitudinal polarization.  In particular, at the $Z$ peak, where
photon effects can be neglected, to go from the unpolarized to
the polarized one, one has to make the replacement (henceforth,
we drop the superscript $Z$, as we shall only deal with $Z$
couplings):
\begin{equation}
\begin{array}{clclc}
g^2_{Ve} + g^2_{Ae} &\rightarrow &g^2_{Ve} + g^2_{Ae} &-& P_e \,2
g_{Ve} g_{Ae},  \\
2g_{Ve}g_{Ae} &\rightarrow &2g_{Ve}g_{Ae} &-
&P_e \left(g^2_{Ve} + g^2_{Ae}\right).
\end{array}
\end{equation}
We have set $P_{\bar e}=0$, as is the case at SLC, for example.
It is then clear from (9), using $\sin^2\theta_W \approx 0.23$,
that quantities which are suppressed in the absence of
polarization because of the small numerical value of
$2g_{Ve}g_{Ae}$ will get considerably enhanced in the presence
of polarization.

To calculate correlations of $O_1$ and $O_2$, we need the
differential cross section for (1) followed by (2) at the $Z$
peak, arising from SM $Z$ couplings of $e$ and $\tau$, together
with a weak-dipole coupling of $\tau$ arising from eq.(4).  (We
neglect electromagnetic effects completely).  The calculation may be
conveniently done, following ref.[3] (see also ref.[13]), in
steps, by first determining the production matrix $\chi$ for
$\tau^+\tau^-$ in spin space, and then taking
its trace with the decay matrices $\cal{D}^{\pm}$ for $\tau^{\pm}$
decays into single charged particle in addition to the
invisible neutrino.

\begin{equation}
\frac{1}{\sigma}\,\frac{d\sigma}{d\Omega_kd\Omega^*_-d\Omega^*_+dE^*_-dE^*_+} =
\frac{k}{8\pi m^2_Z\Gamma(Z\rightarrow\tau^+\tau^-)}
\frac{1}{(4\pi)^3}
\chi^{\beta\beta^{\prime}, \alpha\alpha^{\prime}}
{\cal D}^-_{\alpha^\prime\alpha} {\cal D}^+_{\beta^\prime\beta},
\end{equation}
where $d\Omega_k$ is the solid angle element for
${\bf k}_+$ in the overall c.m. frame, $k =
\vert{\bf k}_+\vert$, and $d\Omega^*_{\pm}$ are
the solid angle elements for ${\bf q}^*_{{\bar B},A}$,
the $\bar{B}$ and $A$
momenta in the $\tau^{\pm}$ rest frame.  The
${\cal D}$ matrices are given by
\begin{eqnarray}
& {\cal D}^{+} = \delta\left(E^*_{B} - E_{0B}\right)\left[ 1 -
\alpha_B {\bf \sigma}_{+} \cdot\hat{\bf q}^*_{B} \right] & \nonumber \\
& {\cal D}^{-} = \delta\left(E^*_{A} - E_{0A}\right)\left[ 1 +
\alpha_A {\bf \sigma}_{-} \cdot\hat{\bf q}^*_{A} \right], &
\end{eqnarray}
where ${\bf \sigma}_{\pm}$ are the Pauli
matrices corresponding to the $\tau^{\pm}$ spin, $E^*_{\pm}$ are
the charged particle energies in the $\tau^{\pm}$ rest frame, and
\begin{equation}
E_{0A,B} = \frac{1}{2} m_{\tau} (1 + p_{A,B}) ;\;
p_{A,B} = m^2_{A,B}/m^2_{\tau}.
\end{equation}

The expressions for $\chi$ arising from SM as well as WDFF
coupling of $\tau$ are rather long, and we refer the reader to
ref.[3] for these expressions in the absence of polarization.
It is straightforward to incorporate polarization using (11).

\vskip 1cm
\noindent{\bf 3. Results}
\vskip 1cm

Using eqns. (13)-(15) above, as well as the expression for the
$\tau^+\tau^-$ production matrix $\chi$ from [3], we can obtain
expressions for $\langle O_1 \rangle$ and $\langle O_2\rangle$
by writing $O_1$ and $O_2$ in terms of the $\tau$ rest frame
variables and carrying out integrals over them analytically. The
expressions for the correlations $\langle O_1\rangle $ and
$\langle O_2\rangle$ obtained
are, neglecting $\tilde{d}^2_{\tau}$,
\begin{eqnarray}\nonumber
& \langle O_1\rangle   =  -
\frac{m^2_Z}{18e}m_{\tau}{\rm Re}\,\tilde{d}_{\tau}
c_Ws_W(1-x^2)\left(\frac{r - P_e}{1-rP_e}\right) &  \nonumber \\
 & \frac{g_{A\tau}\alpha_A \alpha_B(1-p_A)(1-p_B) -
\frac{3}{2}g_{V\tau}[\alpha_A(1-p_A)(1+p_B)+
\alpha_B(1-p_B)(1+p_A)]}{g^2_{V\tau} (1+\frac{1}{2}x^2) +
g^2_{A \tau}
\left( 1-x^2\right)} &,
\end{eqnarray}
and
\begin{eqnarray}
& \langle O_2\rangle = \frac{2m_Z}{3e}
m_{\tau}{\rm Im}\,\tilde{d}_{\tau}c_Ws_W
\left(\frac{r - P_e}{1-rP_e}\right) & \nonumber \\
& \frac{g_{A\tau}(1-x^2)^2
\frac{1}{2}(\alpha_A(1-p_A)
+\alpha_B(1-p_B))}{g^2_{V\tau}(1+\frac{1}{2}x^2) + g^2_{A\tau} (1-x^2)}  ,&
\end{eqnarray}
where $c_W=\cos\theta_W$, $s_W=\sin\theta_W$, $x =
2m_{\tau}/m_Z$, and $r=2g_{Ve}g_{Ae}/(g^2_{Ve}+g^2_{Ae})$.

We have also obtained analytic expressions for the variance
$\langle  O^2\rangle  - \langle  O\rangle ^2 \approx \langle  O^2\rangle $
in each case, arising from the CP-invariant SM part of the
interaction:

\begin{eqnarray}\nonumber
  &   \langle O_1^2\rangle  =
\frac{3}{64}\frac{1}{(g_{V_\tau}^2(1+\frac{1}{2}x^2)
+g_{A_\tau}^2(1-x^2))}m_Z^2 m_\tau^2 &  \nonumber \\
& \biggl(
  \frac{1}{270}(1-p_A)^2(1-p_B)^2
  [g_{V_\tau}^2(12+16x^2+2x^4)+
g_{A_\tau}^2(12-4x^2-8x^4)] &  \nonumber \\
  & +\frac{(1-x^2)}{540}
\left([(1+p_A)^2(1-p_B)^2+(1+p_B)^2(1-p_A)^2] \right. & \nonumber \\
& \left. [12g_{V_\tau}^2(3+4x^2))+36g_{A_\tau}^2(1-x^2)
] \right. & \nonumber \\
& \left. +16\alpha_A\alpha_B
(1-p_B^2)(1-p_A^2)(1-x^2)[g_{V_\tau}^2
-g_{A_\tau}^2]\right)   & \nonumber \\
  & +\frac{4}{45}(1-p_A)(1-p_B)g_{V_\tau}g_{A_\tau}(1-x^2)(1-\frac
{x^2}{6}) & \nonumber \\
 & [\alpha_A (1+p_A)(1-p_B)+\alpha_B (1+p_B)(1-p_A)]
\biggr) &
\end{eqnarray}

\begin{eqnarray}
  & \langle O_2^2\rangle =
\frac{1}{12}\frac{1}{(g_{V_\tau}^2(1+\frac{1}{2}x^2)+g_{A_\tau}^2
(1-x^2))}m_Z^2 &   \nonumber \\
  & \biggl[ \biggl( \frac{1}{10}[(1-p_A)^2+(1-p_B)^2][g_{V_\tau}^2
(1+\frac{7}{4}x^2+x^4)+g_{A_\tau}^2(1+\frac{1}{2}x^2-\frac{3}{2}x^4)]
 & \nonumber \\
&  +\frac{1}{15}\alpha_A \alpha_B (1-p_A) (1-p_B)
[g_{V_\tau}^2(-1-\frac{7}{4}x^2-x^4)+g_{A_\tau}^2(-1+2x^2-x^4)]\biggr)
  & \nonumber \\
  & +\frac{9}{4}
\biggl( \frac{2}{15}(1-x^2)(p_A-p_B)^2[g_{V_\tau}^2(1+\frac{x^2}{4})
+g_{A_\tau}^2(1-x^2)] &  \nonumber \\
  & -\frac{2}{45}g_{V_\tau}g_{A_\tau}(1-x^2)^2(4+x^2)(p_A-p_B)
[\alpha_A(1-p_A)-\alpha_B(1-p_B)]\biggr)\biggr] &
\end{eqnarray}

The results for the significant two-body
decay channels are presented in the tables.  In Tables 1
and 2 we have presented, as in  ref.[3], the values of
$c_{AB}$ for $O_1$ and $O_2$ respectively, defined as
 the correlation for a value of ${\rm Re}\,\tilde{d}_{\tau}$ or
${\rm Im}\,\tilde{d}_{\tau}$ (as the case may be) equal to
$e/m_Z$, for various values of $P_e$.
We have also presented the value of $\sqrt{\langle O^2\rangle }$ and
$\delta\tilde{d}_{\tau}$, which represents the 1 s.d.
upper limit unit on $\tilde{d}_{\tau}$ which can be
placed with a certain sample of events, for 50,000 $Z$'s currently
seen at SLC with 62\% polarization, and for $10^6\, Z$'s,
eventually hoped to be achieved. This 1 s.d. limit is the value
of $\tilde{d}_{\tau}$ which gives a mean value of $O_i$
equal to the s.d. $\sqrt{\langle O^2_i\rangle /N_{AB}}$ in each case:
\begin{equation}
c_{AB}\delta\tilde{d}_\tau=\frac{e}{m_Z}\frac{1}{\sqrt{N_{AB}}}
\sqrt{\langle O^2_i\rangle}.
\end{equation}
Here $N_{AB}$ is the number of events in the channel $A\bar{B}$
(or $\bar{A}B$), and is given by
\begin{equation}
N_{AB}=N_ZB(Z\rightarrow\tau^+\tau^-)B(\tau^-\rightarrow
A\nu_\tau )B(\tau^+\rightarrow \bar{B}\bar{\nu}_\tau).
\end{equation}

As can be seen from Tables 1 and 2, $\langle O_1\rangle $ and
$\langle O_2\rangle $ can probe
respectively ${\rm Re}\,\tilde{d}_{\tau}$ and
${\rm Im}\,\tilde{d}_{\tau}$ down to about $1.4\times 10^{-16} e$ cm
and $4.1\times 10^{-16} e$ cm with the current data from SLC, where
we have added the sensitivities by inverse quadrature to
obtain these sensitivities.  These
limits can be improved by a factor of 5 in future runs of SLC.

These limits can be improved by looking at
correlations of the same observables, but in a sample obtained
by counting the difference between the number of events for a certain
polarization, and for the corresponding sign-flipped polarization.
If the partial cross section for the process for an
$e^-$ polarization $P_e$ is given by
\begin{equation}
d\sigma(P_e)=A + P_e\cdot B,
\end{equation}
then one calculates average values over the distribution given
by the difference
\begin{equation}
d\sigma (P_e) - d\sigma (-P_e) =2 P_e B.
\end{equation}
The correlations are considerably larger, whereas the variances
are unchanged. The event sample is however smaller (by a factor
of about .16$P_e$, which is the left-right asymmetry times degree
of polarization), leading to a larger statistical error. The
sensitivity is nevertheless improved, as can be seen from Table
3. There we present, for the two variables $O_1$ and $O_2$,
$c_{AB}$ (defined as before) for the polarization asymmetrized
sample described above,
and the corresponding  quantities $\sqrt{\langle  O_i^2 \rangle }$
and the 1 s.d. limits on ${\rm Re}\,\tilde{d}_\tau$ and
${\rm Im}\,\tilde{d}_\tau$, respectively, both called
$\delta\tilde{d}_\tau$ in this table for convenience. We give
the limits for two
luminosities, but only for one polarization, $P_e=.62$. Only the
last column changes with $P_e$, and the
improvement in going to $P_e=.75$ is marginal.

 As seen from Table 3, the current
data can yield upper limits of about 5.0$\times 10^{-17} e$ cm for
${\rm Re}\,\tilde{d}_{\tau}$ and 1.5$\times 10^{-16} e$ cm for
${\rm Im}\,\tilde{d}_{\tau}$, where we have again combined
errors from different channels by adding inverse squares. An
integrated luminosity corresponding to $10^6$ Z's
at SLC can improve the limits to about 1.2$\times 10^{-17} e$ cm
and 3.3$\times 10^{-17} e$ cm, respectively. These
should be compared with the 95\% C.L.
ALEPH limit of ${\rm Re}\,\tilde{d}_{\tau}$ $< 3.7\times 10^{-17} e$ cm
obtained by looking for tensor correlations from a sample of
about 650,000 Z's
at LEP, and including several decay channels of $\tau$ [4]. The
disarming simplicity of the correlations considered is reflected
in the fact that analytic expressions presented above yield
results comparable to those from more involved correlations
involving expensive numerical methods and the necessity of
including several decays. The simple two-body channels
considered here more or less set the scale of the limits of
discover since they have larger sensitivity than the leptonic
and three-body channels. Further the expressions given here
would serve as a normalization tool for a Monte Carlo package
should it be constructed.

The real advantage of polarization can be seen in the
sensitivity for the measurement of ${\rm Im}\tilde{d}_{\tau}$.
The limit obtainable at LEP under ideal
experimental conditions from the 2$\pi$ channel using the tensor
correlation $\langle
\hat{\bf p}\cdot (\hat{\bf q}_++\hat{\bf q}_-)\hat{\bf p}\cdot
(\hat{\bf q}_+ -\hat{\bf q}_- )\rangle $ is $10^{-16}$ $e$ cm
with $10^7$ $Z$'s [3], whereas the limit we obtain here in the
presence of polarization is as low as
$5\times 10^{-17}$ $e$ cm with only $10^6$ $Z$'s for the same channel.

\vskip 1cm
\noindent{\bf 4. Conclusions}
\vskip 1cm

To conclude, our results show that the CP-odd correlations
$\langle O_1\rangle $
and $\langle O_2\rangle $ are considerably enhanced due to longitudinal $e^-$
polarization.  By considering  values of these
correlations on changing the sign of the polarization, a much
greater sensitivity is obtained. Considering both these
correlations together can give phase information on the WDFF,
which cannot be obtained using any single correlation.
While the sensitivity obtainable at SLC for the real part of the WDFF
is comparable to that obtained at LEP, the sensitivity for the
imaginary part is better than that obtainable at LEP with an
ordinary of magnitude lower luminosity at SLC.

We have neglected errors in measurement of $P_e$,
whose effect on the results would be small compared to that of
statistical errors. We have also ignored radiative corrections,
which may alter our numerical estimates somewhat, and must be
taken into account in a more complete analysis.

It is possible that longitudinal polarization may be available
at tau-charm factories or other future linear colliders planned
to operate at higher
energies.  The advantages of polarization presented here can be
studied by extending the analysis to the relevant energies,
including the effect of photon exchange and the $\tau$ electric
dipole coupling.  The study could also be extended to dipole
moments of the top-quark
or $W^{\pm}$  which could be pair produced at the
high energy accelerators.
The more complete study presented here will serve as
a useful cross-check on Monte-Carlos used in the evaluation
of the correlations and their variances.  Further, in the
event of a very massive sequential lepton found say at the
JLC which is likely to have several two-body decay modes,
the expressions given here can be easily modified to
evaluate the CP-violating form factors of such a lepton,
albeit with the interference from an electric-dipole form
factor since the Z-cross section would no longer dominate.

It would be worthwhile to calculate order-$\alpha$ corrections
to the results obtained above. However, this would be
model dependent, and requires considerable care and effort.

It would also be interesting to consider event asymmetries
corresponding to $O_1$ and $O_2$ (rather than their mean values), and
also other CP-odd
correlations with an odd number of $\hat{\bf p}$'s, like
$\langle  ({\bf q}^2_+ -
{\bf q}^2_- )
\hat{\bf p}\cdot({\bf q}_+ -
{\bf q}_-)\rangle $, which may have a different
sensitivity.
\vskip .2cm
\centerline{\bf Acknowledgements}

B. A. thanks Prof. G. Wanders for a conversation.
Support from the Swiss National Science Foundation during the
course of this work for B. A. is gratefully acknowledged.
S.D.R. thanks D. Indumathi, Anjan Joshipura, Marek Nowakowski
and B.R. Sitaram for discussions. He also thanks Prof. L.M.
Sehgal for discussions and drawing his attention to refs. [11]
and [12].
\vskip 1cm

\newpage
\begin{center}
{\large \bf References}
\end{center}

\noindent [1]  J.F. Donoghue and G. Valencia, Phys. Rev. Lett. {\bf 58}, 451
(1987); M. Nowakowski and A. Pilaftsis, Mod. Phys. Lett. A {\bf
4}, 829 (1989), Z. Phys. C {\bf 42}, 449
(1989); W. Bernreuther, U. L\" ow, J.P. Ma and O. Nachtmann,
Z. Phys. C {\bf 43}, 117 (1989); W.
Bernreuther and O. Nachtmann, Phys. Lett. B {\bf 268}, 424
(1991); M.B. Gavela {\it et al.}, Phys. Rev. D {\bf 39}, 1870 (1989).

\noindent [2] F. Hoogeveen and L. Stodolsky, Phys. Lett. B {\bf 212},
505 (1988); S. Goozovat and C.A. Nelson, Phys. Lett. B {\bf
267}, 128 (1991); G. Couture, Phys. Lett. B {\bf 272}, 404
(1991); W. Bernreuther and O. Nachtmann, Phys. Rev. Lett. {\bf
63}, 2787 (1989).

\noindent [3] W. Bernreuther, G.W. Botz, O. Nachtmann and P. Overmann,
Z. Phys. C {\bf 52}, 567 (1991); W.
Bernreuther, O. Nachtmann and P. Overmann, Phys. Rev. D {\bf
48}, 78 (1993).

\noindent [4] OPAL Collaboration, P.D. Acton {\it et al.}, Phys. Lett. B {\bf
281}, 405 (1992); ALEPH Collaboration, D. Buskulic {\it et al.}, Phys.
Lett. B {\bf 297}, 459 (1992).

\noindent [5] M. Swartz, private communication.

\noindent [6] SLD Collaboration, K. Abe {\it et al.}, Phys. Rev. Lett. {\bf
70}, 2515 (1993).

\noindent [7] M.J. Fero, SLAC report SLAC-PUB-6027 (1992), to be published.

\noindent[8] B. Ananthanarayan and S. D. Rindani,
Physical Research Laboratory preprint, PRL-TH-93/17

\noindent[9] K. Hagiwara, A. D. Martin and D. Zeppenfeld,
Phys. Lett. B {\bf 235}, 198 (1990); See also W. Bernreuther,
O. Nachtmann and P. Overmann in Ref. [3].

\noindent [10] A similar enhancement for the observation of CP violation
in B decays was pointed out by W.B. Atwood, I. Dunietz and P.
Grosse-Wiesmann, Phys. Lett. B {\bf 216}, 227 (1989).

\noindent [11] T.D. Lee and M. Nauenberg, Phys. Rev. {\bf B
133}, 1549 (1964); F. Berends {\it et al.}, Nucl.
Phys. {\bf B239}, 382,395 (1984); H.F. Contopanangos and M.B.
Einhorn, Nucl. Phys. {\bf B377}, 20 (1992).

\noindent [12] B. Falk and L.M. Sehgal, Aachen report no. PITHA
93/29 (Rev.), to appear in Phys. Lett. B.

\noindent [13] Y.S. Tsai, Phys. Rev. {\bf D4}, 2821 (1971); {\bf 13},
771 (1976) (E); S. Kawasaki, T. Shirafuji and Y.S. Tsai, Prog.
Theo. Phys. {\bf 49}, 1656 (1973).

\newpage

\noindent{\bf Tables}
\vskip 1cm
\noindent 1. Correlations and 1 s.d. limits on Re
$\tilde{d}_\tau$ for the observable $O_1$ corresponding to the
final states (a)$\pi\pi$ (b)$\pi\rho$ and (c)$\rho\rho$.
\vskip 1cm
\noindent 2. Correlations and 1 s.d. limits on Im
$\tilde{d}_\tau$ for the observable $O_2$ corresponding to the
final states (a)$\pi\pi$ (b)$\pi\rho$ and (c)$\rho\rho$.
\vskip 1cm
\noindent 3. Results for the polarization asymmetrized sample of
events for $P_e=0.62.$ Here $\delta \tilde{d}_\tau$ refers to
its real part in case of $O_1$ and to its imaginary part for
$O_2$.
\begin{center}
\begin{tabular}{||c|c|c|c|c||}
\hline
$P_e$ &$c_{\pi\pi}$ &
$\sqrt{\langle  O^2_1\rangle }$ &\multicolumn{2}{|c||}{$\delta
{\rm Re}\,\tilde{d}_{\tau}(e$ cm) for}\\
\cline{4-5}
 &(GeV$^2$) &(GeV$^2$)&
{$5\times 10^4\, Z$'s} &{$10^6
\, Z$'s}\\
\hline
0 &0.898 &$12.861$ &$6.6\times 10^{-16}$ &$1.5\times 10^{-16}$\\
$+0.62$ &$-2.890$ &12.861 &$2.0\times 10^{-16}$ &$4.6\times 10^{-17}$\\
$-0.62$ &$4.007$ &12.861 &$1.5\times 10^{-16}$ &$3.3\times 10^{-17}$\\
$+0.75$ &$-3.792$ &12.861 &$1.6\times 10^{-16}$ &$3.5\times 10^{-17}$\\
$-0.75$ &$4.589$ &12.861 &$1.3\times 10^{-16}$ &$2.9\times 10^{-17}$\\
\hline
\end{tabular}
\vskip 0.5cm
Table 1(a)
\end{center}
\vskip 1cm
\begin{center}
\begin{tabular}{||c|c|c|c|c||}
\hline
$P_e$ &$c_{\pi\rho}$ &
$\sqrt{\langle  O^2_1\rangle }$ &\multicolumn{2}{|c||}{$\delta
{\rm Re}\,\tilde{d}_{\tau}(e$ cm) for}\\
\cline{4-5}
 &(GeV$^2$) &(GeV$^2$)&
{$5\times 10^4\, Z$'s} &{$10^6
\, Z$'s}\\
\hline
0 &0.223 &$13.433$ &$2.0\times 10^{-15}$ &$4.5\times 10^{-16}$\\
$+0.62$ &$-0.716$ &13.433 &$6.2\times 10^{-16}$ &$1.4\times 10^{-16}$\\
$-0.62$ &$0.993$ &13.433 &$4.5\times 10^{-16}$ &$1.0\times 10^{-16}$\\
$+0.75$ &$-0.940$ &13.433 &$4.8\times 10^{-16}$ &$1.1\times 10^{-16}$\\
$-0.75$ &$1.137$ &13.433 &$3.9\times 10^{-16}$ &$8.8\times 10^{-17}$\\
\hline
\end{tabular}
\vskip 0.5cm
Table 1(b)
\vskip 1cm
\end{center}
\begin{center}
\begin{tabular}{||c|c|c|c|c||}
\hline
$P_e$ &$c_{\rho\rho}$ &
$\sqrt{\langle  O^2_1\rangle }$ &\multicolumn{2}{|c||}{$\delta
{\rm Re}\,\tilde{d}_{\tau}(e$ cm) for}\\
\cline{4-5}
 &(GeV$^2$) &(GeV$^2$)&
{$5\times 10^4\, Z$'s} &{$10^6
\, Z$'s}\\
\hline
0 &0.040 &$12.874$ &$7.7\times 10^{-15}$ &$1.7\times 10^{-15}$\\
$+0.62$ &$-0.130$ &12.874 &$2.4\times 10^{-15}$ &$5.4\times 10^{-16}$\\
$-0.62$ &$0.180$ &12.874 &$1.7\times 10^{-15}$ &$3.9\times 10^{-16}$\\
$+0.75$ &$-0.170$ &12.874 &$1.8\times 10^{-15}$ &$4.1\times 10^{-16}$\\
$-0.75$ &$0.206$ &12.874 &$1.5\times 10^{-15}$ &$3.4\times 10^{-16}$\\
\hline
\end{tabular}
\\ \vskip .5cm
Table 1(c)
\end{center}

\begin{center}
\begin{tabular}{||c|c|c|c|c||}
\hline
$P_e$ &$c_{\pi\pi}$
&$\sqrt{\langle  O^2_2\rangle }$&\multicolumn{2}{c||}{$\delta
{\rm Im}\,\tilde{d}_{\tau} (e$ cm) for} \\
\cline{4-5}
&(GeV) &(GeV)
&$5\times 10^4\, Z$'s
&$10^6\,  Z$'s\\
\hline
0 &$-0.157$ &9.572 &$2.8\times 10^{-15}$ &$6.2\times 10^{-16}$\\
$+0.62$ &$0.505$ &9.572 &$8.7\times 10^{-16}$ &$1.9\times 10^{-16}$\\
$-0.62$ &$-0.700$ &9.572 &$6.3\times 10^{-16}$ &$1.4\times 10^{-16}$\\
$+0.75$ &$0.662$ &9.572 &$6.6\times 10^{-16}$ &$1.5\times 10^{-16}$\\
$-0.75$ &$-0.802$ &9.572 &$5.5\times 10^{-16}$ &$1.2\times 10^{-16}$\\
\hline
\end{tabular}
\vskip 0.5cm
Table 2(a)
\vskip 1cm
\end{center}

\begin{center}
\begin{tabular}{||c|c|c|c|c||}
\hline
$P_e$ &$c_{\pi\rho}$
&$\sqrt{\langle  O^2_2\rangle }$&\multicolumn{2}{c||}{$\delta
{\rm Im}\,\tilde{d}_{\tau} (e$ cm) for} \\
\cline{4-5}
&(GeV) &(GeV)
&$5\times 10^4\, Z$'s
&$10^6\,  Z$'s\\
\hline
0 &$-0.108$ &10.324 &$3.2\times 10^{-15}$ &$7.1\times 10^{-16}$\\
$+0.62$ &$0.347$ &10.324 &$9.9\times 10^{-16}$ &$2.2\times 10^{-16}$\\
$-0.62$ &$-0.482$ &10.324 &$7.1\times 10^{-16}$ &$1.6\times 10^{-16}$\\
$+0.75$ &$0.456$ &10.324 &$7.5\times 10^{-16}$ &$1.7\times 10^{-16}$\\
$-0.75$ &$-0.552$ &10.324 &$6.2\times 10^{-16}$ &$1.4\times 10^{-16}$\\
\hline
\end{tabular}
\vskip 0.5cm
Table 2(b)
\vskip 1cm
\end{center}

\begin{center}
\begin{tabular}{||c|c|c|c|c||}
\hline
$P_e$ &$c_{\rho\rho}$
&$\sqrt{\langle  O^2_2\rangle }$&\multicolumn{2}{c||}{$\delta
{\rm Im}\,\tilde{d}_{\tau} (e$ cm) for} \\
\cline{4-5}
&(GeV) &(GeV)
&$5\times 10^4\, Z$'s
&$10^6\,  Z$'s\\
\hline
0 &$-0.059$ &9.246 &$3.8\times 10^{-15}$ &$8.4\times 10^{-16}$\\
$+0.62$ &$0.190$ &9.246 &$1.2\times 10^{-15}$ &$2.6\times 10^{-16}$\\
$-0.62$ &$-0.264$ &9.246 &$8.5\times 10^{-16}$ &$1.9\times 10^{-16}$\\
$+0.75$ &$0.249$ &9.246 &$8.9\times 10^{-16}$ &$2.0\times 10^{-16}$\\
$-0.75$ &$-0.302$ &9.246 &$7.4\times 10^{-16}$ &$1.7\times 10^{-16}$\\
\hline
\end{tabular}
\\ \vskip .5cm
Table 2(c)
\end{center}

\begin{center}
\begin{tabular}{||c|c|c|c|c||}
\hline
 &$c_{\pi\pi}$ &$\sqrt{\langle  O^2_i\rangle }$ &\multicolumn{2}{c||}{$\delta
\tilde{d}_\tau (e\;{\rm cm})$ for} \\
\cline{4-5}
& & &$5\times 10^4\, Z$'s
&$10^6\,  Z$'s\\
\hline
$O_1$ &35.545 GeV$^2$&12.861 GeV$^2$ &$5.3\times 10^{-17}$ & $1.2\times
10^{-17}$\\
$O_2$ &$-6.208$ GeV&$9.572$ GeV &$2.2\times 10^{-16}$ &$5.0\times 10^{-17}$\\
\hline
\end{tabular}
\vskip 0.5cm
Table 3(a)
\vskip 1cm
\end{center}

\begin{center}
\begin{tabular}{||c|c|c|c|c||}
\hline
 &$c_{\pi\rho}$ &$\sqrt{\langle  O^2_i\rangle }$ &\multicolumn{2}{c||}{$\delta
\tilde{d}_\tau (e\;{\rm cm})$ for} \\
\cline{4-5}
& & &$5\times 10^4\, Z$'s
&$10^6\,  Z$'s\\
\hline
$O_1$ &8.810 GeV$^2$&13.433 GeV$^2$ &$1.6\times 10^{-16}$ & $3.6\times
10^{-17}$\\
$O_2$ &$-4.273$ GeV&$10.324$ GeV &$2.6\times 10^{-16}$ &$5.7\times 10^{-17}$\\
\hline
\end{tabular}
\vskip 0.5cm
Table 3(b)
\vskip 1cm
\end{center}

\begin{center}
\begin{tabular}{||c|c|c|c|c||}
\hline
 &$c_{\rho\rho}$ &$\sqrt{\langle  O^2_i\rangle }$ &\multicolumn{2}{c||}{$\delta
\tilde{d}_\tau (e\;{\rm cm})$ for} \\
\cline{4-5}
& & &$5\times 10^4\, Z$'s
&$10^6\,  Z$'s\\
\hline
$O_1$ &1.594 GeV$^2$&12.874 GeV$^2$ &$6.2\times 10^{-16}$ &
$1.4\times 10^{-16}$\\
$O_2$ &$-2.338$ GeV&$9.246$ GeV &$3.0\times 10^{-16}$
&$6.8\times 10^{-17}$\\
\hline
\end{tabular}
\\ \vskip .5cm
Table 3(c)
\end{center}

\end{document}